\documentstyle[epsf,rotate,twocolumn,prl,aps,rotate]{revtex}

\begin{document}

\twocolumn[\hsize\textwidth\columnwidth\hsize\csname @twocolumnfalse\endcsname

\title{
\vspace*{-12mm}{\normalsize\sl
\hspace{-4cm}To appear in the
Proceedings of {\it Spectroscopies of Novel
Superconductors, Cape Cod, September 14-18, 1997}}\\
\vspace*{10mm}
Andreev Bound States, Surfaces and Subdominant Pairing\\
in High T$_c$ Superconductors
}
\author{ D. Rainer$^{a}$, 
H. Burkhardt$^{a}$, M. Fogelstr\"om$^{b}$, and J. A. Sauls$^{b}$}
\address{
$^a$Physikalisches Institut, Universit\"at Bayreuth, D-95440 Bayreuth, 
Germany\\
$^b$Department of Physics \& Astronomy, Northwestern University,
Evanston, IL 60208\\}

\maketitle

\begin{abstract}
A characteristic feature of the BCS theory of 
superconductivity is the quantum-mechanical coherence of 
particle and hole states. Direct observation of particle-hole 
coherence in unusual superconducting materials
is a strong indication of traditional superconductivity. We use the
Fermi liquid theory of superconductivity to study 
the implications of
particle-hole coherence on properties of
d-wave superconductors near surfaces. Typical surface phenomena
are the suppression of the superconducting order parameter, 
surface bound states associated with
Andreev reflection, anomalous screening currents, 
and spontaneous breaking
of time-reversal symmetry. We review these phenomena
and present new results for
the effects of surface roughness.
\end{abstract}
\vspace*{3mm}
] 

It is generally accepted that high T$_c$ cuprate superconductors
are strongly correlated metals whose superconducting state is not
well described by mean field theory or equivalent 
methods designed for weakly correlated electrons.
On the other hand, theoretical
methods for studying strongly correlated electrons,
developed in the context
of  high T$_c$ superconductivity, are
still far from being able to 
calculate subtle effects such as the influence
of surface roughness on 
superconducting properties. A theory 
which is not restricted to weakly correlated systems 
but nevertheless
can be used to study
a broad range of superconducting phenomena including 
the effects of surfaces 
is the Fermi liquid theory of superconductivity.
The most useful
formulation of  this
theory is in terms of 
{\em quasiclassical transport equations}  
derived in 1968 by Eilenberger\cite{eil68} and Larkin
\& Ovchinnikov\cite{lar68}
for superconductors in equilibrium, and  generalized to 
nonequilibrium phenomena
a few years later\cite{eli71,lar76}. 
The Fermi liquid theory of superconductivity
combines the motion of quasiparticles along
classical trajectories ({\em external degrees
of freedom})
with the quantum dynamics
of {\em internal degrees of freedom}
which are the spin
and the particle-hole
degrees of freedom. This
combination of classical and
quantum physics is the 
proper generalization of Landau's
semi-classical transport
equation for normal Fermi liquids 
to the superconducting state. 
Like Landau's theory the 
leading order terms in the expansion
parameters of Fermi liquid theory
are low-energy compared with the Fermi
energy or long wavelength compared
with the atomic scale, e.g. $1/k_f\xi_0$, where
$\xi_0=\hbar v_f/2\pi T_c$\cite{rai95,rai95a}.
The equations of the
Fermi liquid theory of superconductivity
consist of the quasiclassical  transport equation for the
quasiclassical propagator, $\check g(\vec{p}_f,\vec R; \epsilon,t)$,
Eilenberger's normalization condition
for $\check{g}$, the self-consistency equations for the
self-energies, $\check{\sigma}(\vec{p}_f,\vec R; \epsilon,t)$, and
boundary conditions at surfaces and interfaces\cite{ovc69,ser83,kur87,ash89}.
An important part of the self-energy is the superconducting order parameter
$\check{\Delta}$; it establishes the coherence between particles and holes.
We refer to recent reviews \cite{rai95,rai95a} and the
original publications \cite{eil68,lar68,eli71} 
for the definitions and physical interpretation of the
quasiclassical propagators and self-energies, and the detailed 
form of the quasiclassical transport equations and boundary conditions.

In quasiclassical theory an excitation   approaches the surface 
along a classical (straight) incoming trajectory
and is reflected into an
outgoing trajectory. At a specular surface 
the outgoing trajectory is fixed by 
the  conservation 
of parallel momentum (ideal reflection).  
Surfaces with roughness lead to  
a statistical distribution of outgoing trajectories.
This classical picture  for the kinematics of 
an excitation  must be
supplemented by the quantum equations for the
internal degrees of freedom. 
The internal state
along a classical trajectory is obtained
by solving the quasiclassical 
matrix transport equations on this trajectory.
The most important quantum effect  in this 
context is Andreev reflection\cite{and64}, which is caused by 
rotations of the internal  state of an excitation 
from {\em particle-type} to {\em hole-type} (or vice versa).
This may lead to a velocity reversal (retro-reflection) or to 
trapping of an excitation (Andreev bound states).
Particle-hole rotation  and Andreev reflection
are controlled by the off-diagonal self-energy (order parameter), 
and carry information about the anisotropy and
symmetry of the order parameter. 
A typical trajectory at a (120) surface of a d$_{x^2-y^2}$ 
superconductor is shown in Fig.1. For this trajectory
repeated Andreev reflections 
lead to a bound state with 
excitation energy $\epsilon=0$, a zero-energy bound state (ZBS),
which is a robust feature of the spectrum depending only on
the change in sign of the order parameter
along the trajectory\cite{ati75}.

A quasiclassical study of  surface effects in
unconventionally paired superfluids (e.g. p-wave pairing in $^3$He)
was published by Ambegaokar et al.\cite{amb74},
who used deGennes' method of classical
correlation functions\cite{deg64}
to study the suppression and reorientation of the order parameter. 
DeGennes' method
is a predecessor of the full quasiclassical theory applicable to the 
limit $\Delta\!\rightarrow\! 0$. The order
parameter and the tunneling spectra at ideal surfaces of a p-wave 
superconductor in the Balian-Werthamer state
were calculated  by Buchholtz and Zwicknagl\cite{buc81}.
These authors found surface bound states and, in
particular, a ZBS for trajectories at perpendicular incidence;
for the Balian-Werthamer state the conditions of the 
Atiyah-Patodi-Singer theorem are fulfilled 
only at perpendicular incidence. The ZBS  shifts
for all other angles of incidence. \\[-42pt]

\begin{figure}[t]
\vspace*{10mm}
\begin{center}
\begin{minipage}[b]{50mm}
\epsfysize=33mm {\epsfbox{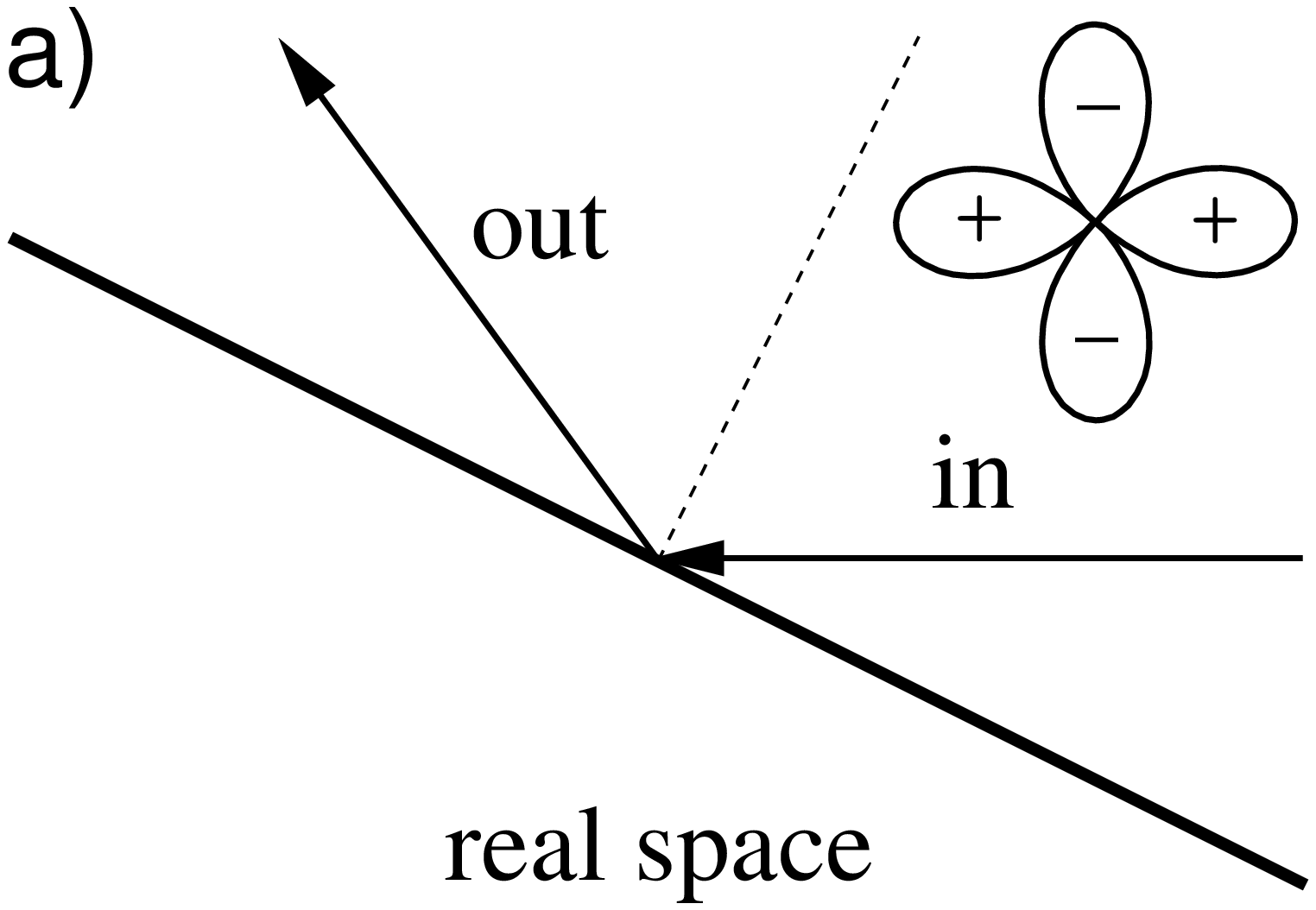}}
\vspace*{8mm}
\end{minipage}\\

\begin{minipage}[b]{50mm}
\epsfysize=33mm {\epsfbox{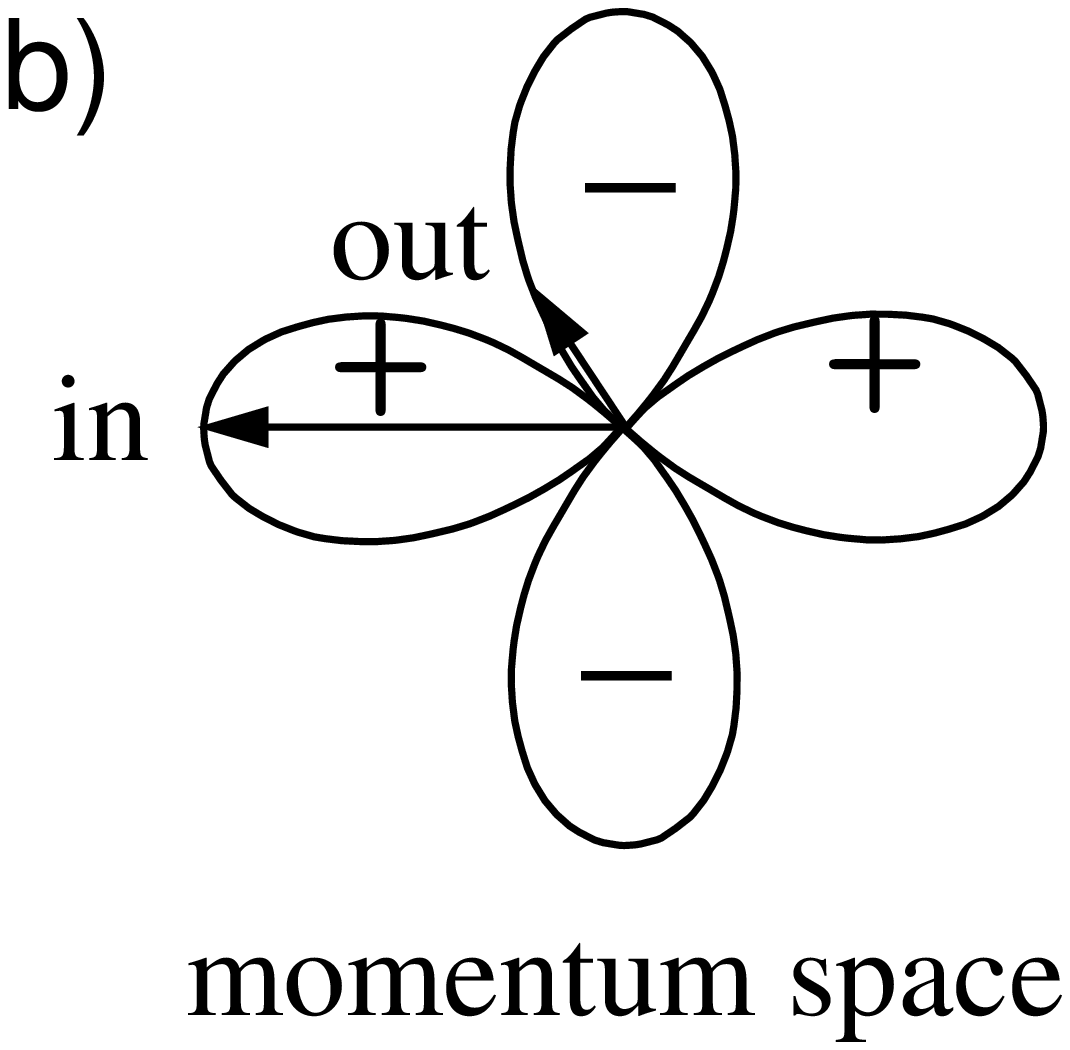}}
\end{minipage}\\[12pt]

\begin{minipage}[b]{50mm}
\epsfxsize=93mm {\epsfbox{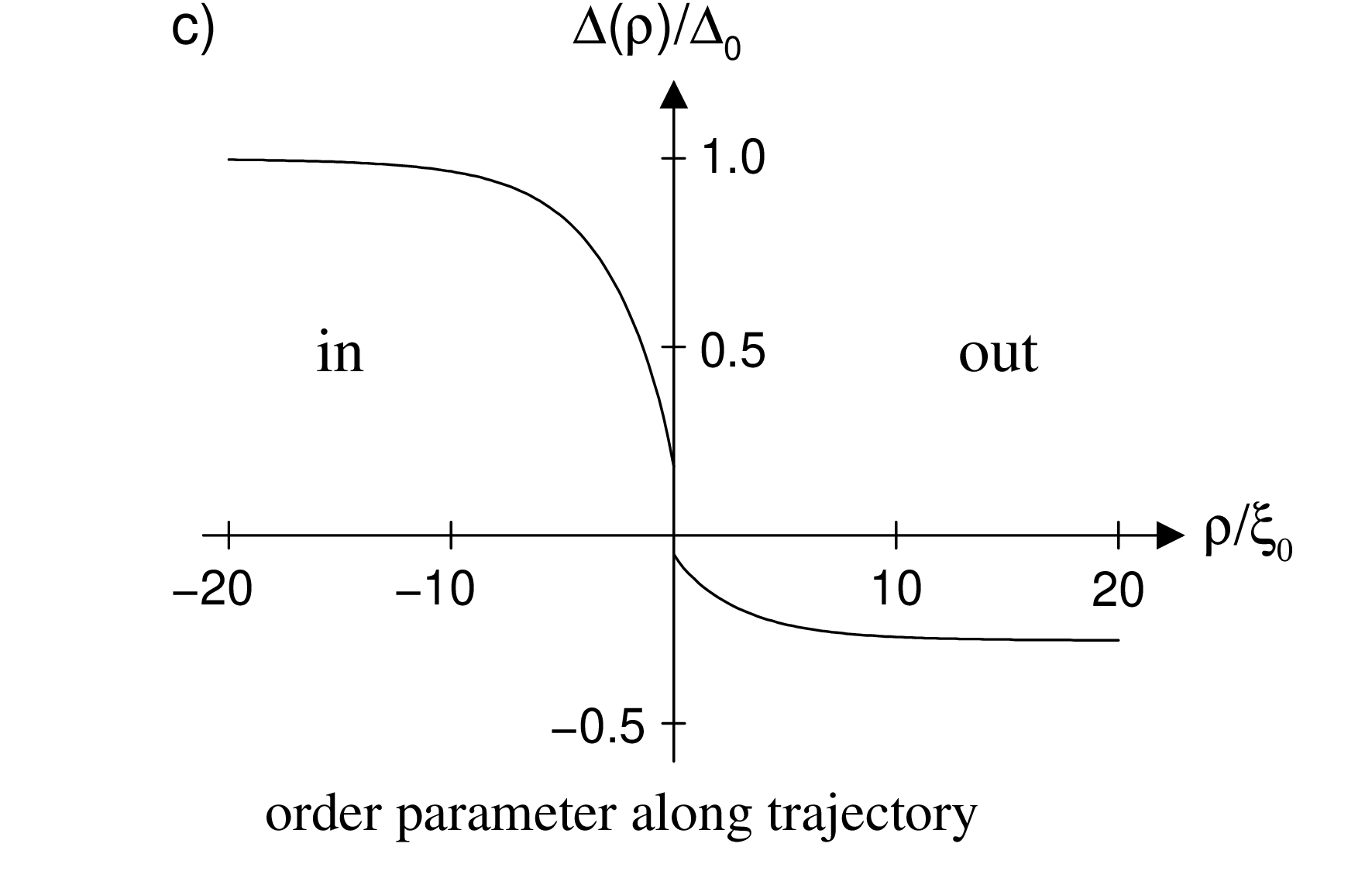}}
\end{minipage}
\end{center}
\caption{Formation of an Andreev bound state at an ideal 
(120) surface.
An excitation  moving along the
trajectory  in Fig.1a experiences changes in the order parameter 
$\Delta(\vec{p}_f,\vec{R})$, 
where $\vec{p}_f$ is the momentum of the excitation,
and $\vec{R}$ is the position. These changes are due to 
1) the depletion of the order parameter near the surface and 2) 
the change in momentum direction of the excitation 
when hitting the surface. The change in
momentum direction leads to a sign change of the order parameter
as shown in Fig.1b. Fig.1c shows a sketch
of the order parameter along the trajectory of the excitation.
\label{Fig1}
}
\end{figure}
\vspace{-12pt}
The chances  for observing zero-energy bound states
are more favorable for d-wave pairing\cite{hu94}.
In this case the ZBS exists at 
an ideal surface for a whole range of angles.
For  (110) surfaces and  $d_{x^2-y^2}$ pairing in bulk 
this range covers  all angles of incidence, while the range
collapses to zero at the 
(100) and (010) surfaces. These results were confirmed
and extended by 
several authors\cite{tan95,mat95,buc95}.
The ZBS leads to a zero-bias conductance peak which  was 
observed in tunneling experiments on cuprate high-T$_c$ 
superconductors by several groups\cite{gee88,les92,kas94,ric96,cov96}.
The zero-bias conductance peak has been identified 
convincingly as due to tunneling into the ZBS
\cite{cov97,alf97,fog97}. Hence, the existence of a ZBS associated with
an unconventional d-wave order parameter in 
high-T$_c$ superconductors seems well 
established by experiment as well as theory.
 
The order parameter near T$_c$ is determined
by the solution of the linearized gap-equation at T$_c$.
The symmetry of this solution
defines the dominant pairing channel; subdominant pairing channels may
mix spontaneously in bulk superconductors below a transition temperature
T$_{sub}<\,$T$_{c}$.
However, no such phase transition has been observed 
in high-T$_c$ superconductors. Thus,
either there is no attractive channel other than the dominant one, or the 
subdominant order parameter is blocked in bulk by the dominant order parameter.
In the latter case, a subdominant superconducting
order parameter which is not subject to
 surface pair breaking (s-wave pairing, etc.)
may be stabilized at surfaces where the dominant order parameter is 
suppressed. This possibility has stimulated theoretical
work by Matsumoto and Shiba\cite{mat95} and Buchholtz et al.\cite{buc95} on
the mixing of subdominant pairing channels at surfaces.
Of particular interest is the possibility of time-reversal 
symmetry breaking by the subdominant order parameter, as
first discussed in the framework of Ginzburg-Landau theory 
by Sigrist et al.\cite{sig95} 
This effect was studied in the full temperature range by 
Matsumoto and Shiba\cite{mat95} who used 
the quasiclassical theory to calculate the
structure of the order parameter, the quasiparticle excitation spectrum 
at the surface, and the spontaneous surface currents.
\vspace{-20pt}

\begin{figure}[t]
\vspace*{5mm}
\centerline{
\epsfysize=0.45\textwidth {\rotate[r]{\epsfbox{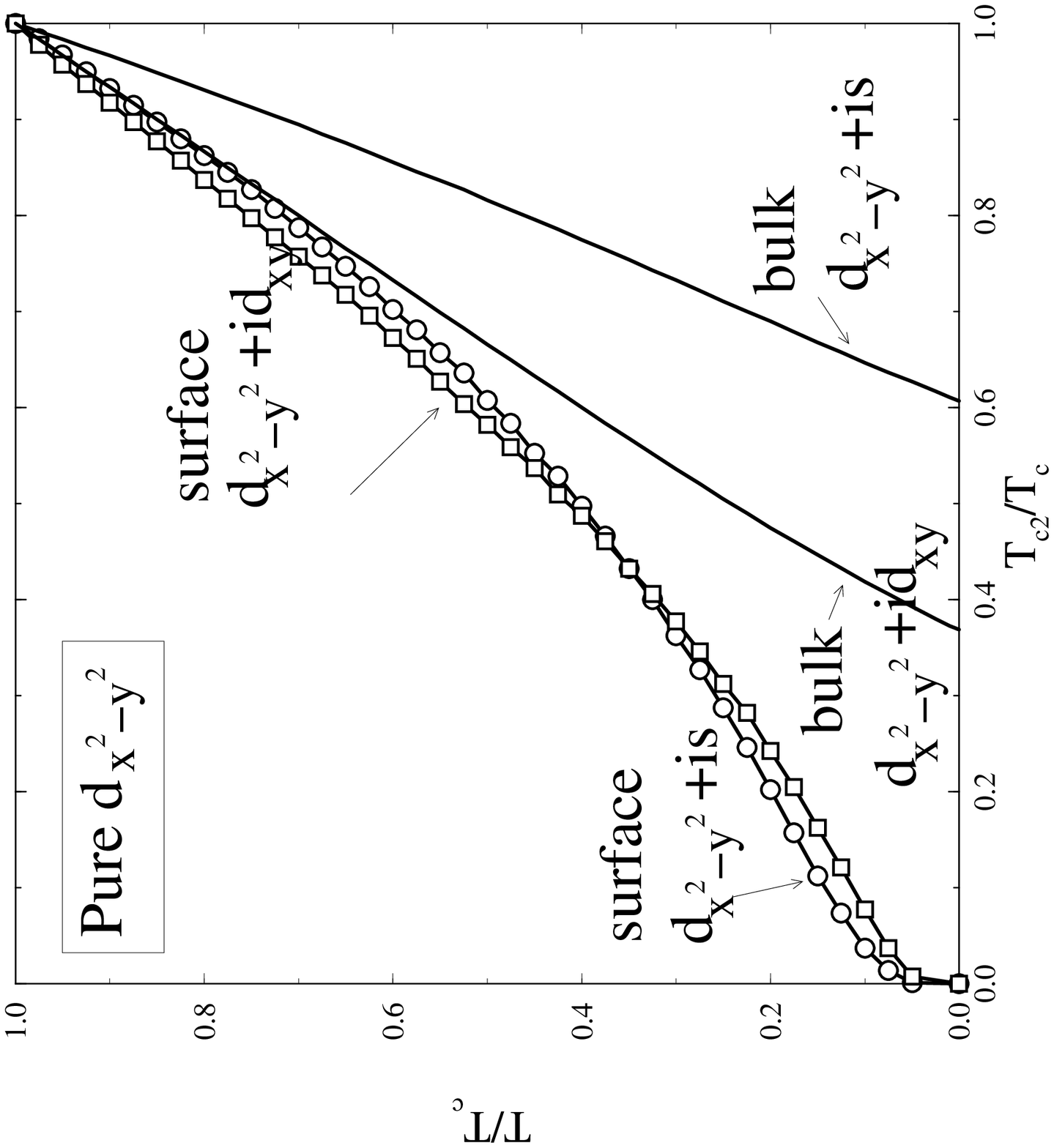}}}
}
\caption{Calculated phase diagram
for the surfaces phases of type 
$d_{x^2-y^2}+ is$ and 
$d_{x^2-y^2}+ id_{xy}$ at an 
ideal (110) surface of a 
d$_{x^2-y^2}$--superconductor
with a bulk transition
temperature $T_{c}$. 
The bulk transition lines are
also shown.
\label{Fig2}
}
\end{figure}
Fig.2 shows
the phase diagram for the transition to a time-reversal breaking surface phase
at the  (110) surface of a d$_{x^2-y^2}$-superconductor,
as calculated by Fogelstr\"om et al.\cite{fog96}.
The transition temperatures of the surface phases are given as a function of
the strength of the subdominant pairing interaction, which we measure by the
subdominant transition temperature  T$_{c2}$.
Also shown are the corresponding bulk transitions. We note that the
order parameters with $s$ or $d_{xy}$ symmetry are not suppressed by
an ideal (110) surface, whereas the bulk stable phase, of type $d_{x^2-y^2}$,
is suppressed by the surface. As a result the sub-dominant order parameter
can nucleate in a region of strong pair-breaking, and the corresponding
surface phase spontaneously breaks time-reversal symmetry for $T<T_{s}$.
\\[-28pt]
\begin{figure}[t]
\vspace*{10mm}
\begin{center}
\begin{minipage}[b]{50mm}
\epsfysize=63mm {\epsfbox{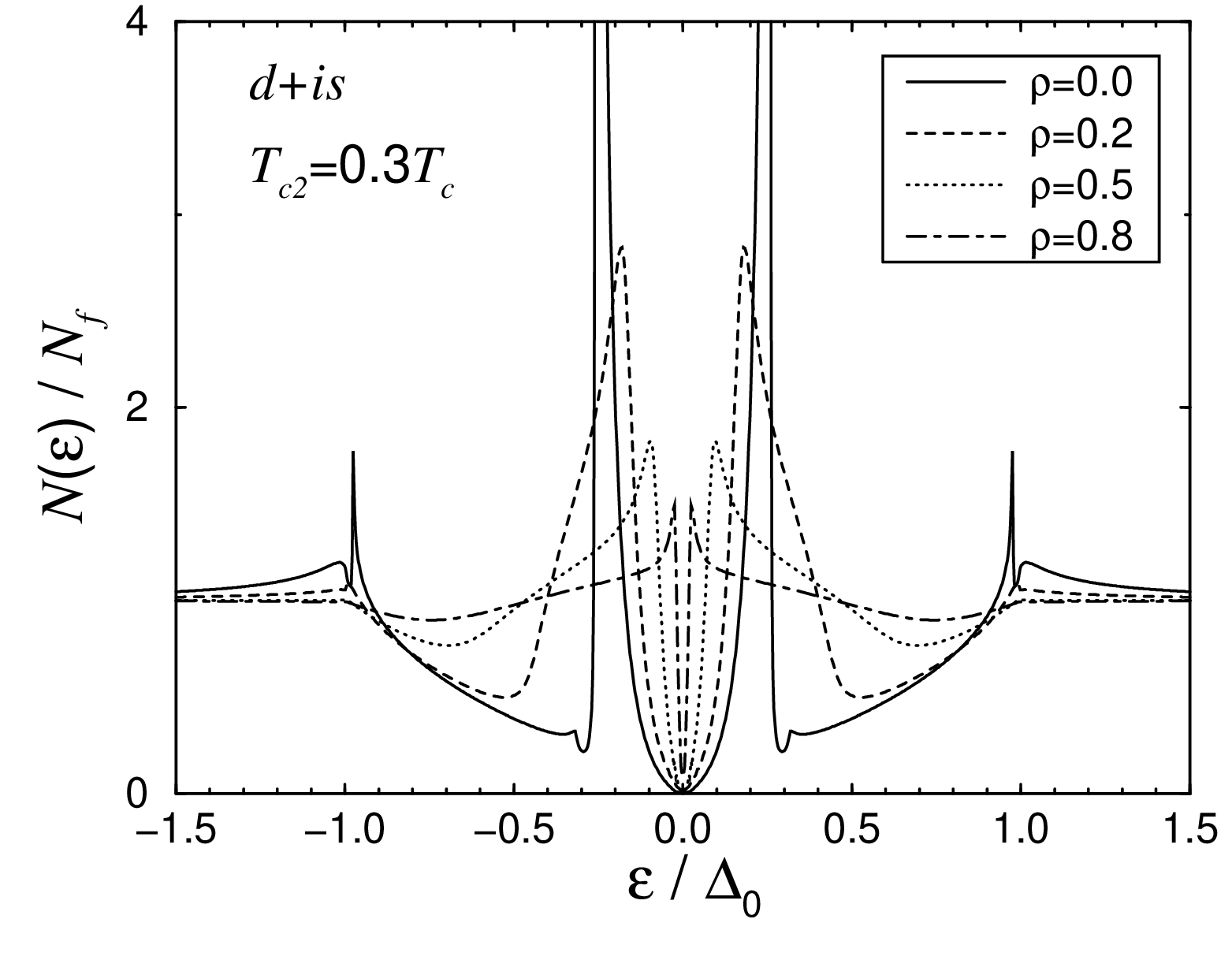}}
\end{minipage}
\\
\begin{minipage}[b]{50mm}
\epsfysize=63mm {\epsfbox{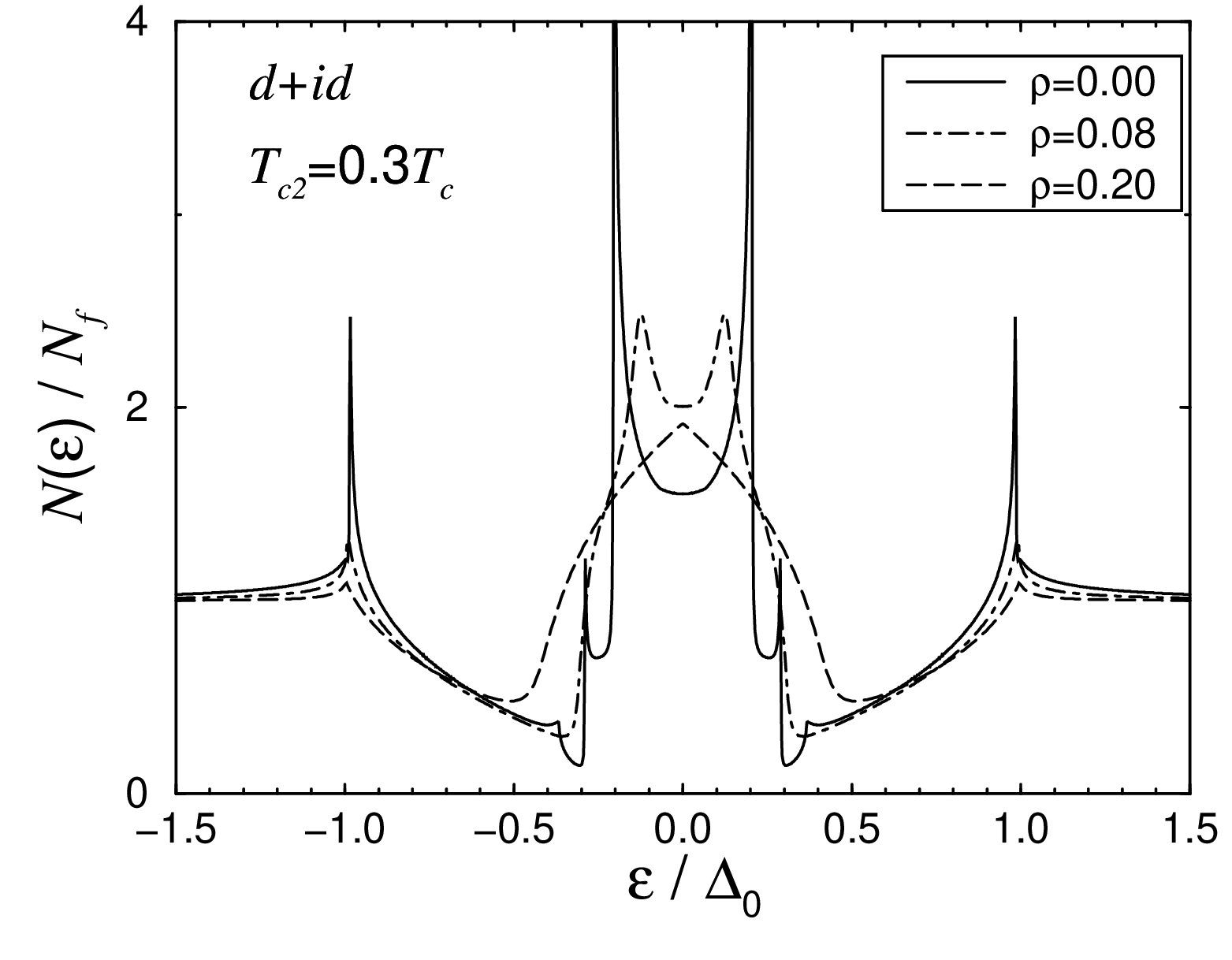}}
\end{minipage}
\end{center}
\vspace*{0mm}
\caption{Tunneling density of states
at a (110) surface
as function of energy ($\epsilon$).
Numerical results are presented for
$T_{c2}/T_{c}=0.3$, $T/T_{c}=0.1$,
various degrees of surface roughness ($\rho$),
and two different 
subdominant order parameters.
\label{Fig3}
}
\end{figure}
The tunneling density of states  at a (110) surface
is shown in Fig.3. This figure demonstrates  the 
characteristic splitting of the ZBS due to
time-reversal breaking surface phases,
and the effects of surface roughness.
Note the presence of two pairs of time-reversed 
Andreev bound states for the $d_{x^2-y^2}+id_{xy}$
surface phase.
Surface roughness is described by Ovchinnikov's
model.\cite{ovc69}
The degree of surface roughness is measured by the  
parameter $\rho$\cite{bar97}, where the ideal
surface corresponds  to  $\rho=0$. 
The state $d_{x^2-y^2}+id_{xy}$ is
clearly more sensitive to surface 
roughness than the state $d_{x^2-y^2}+is$.
An important consequence of  time-reversal breaking 
surface states is a spontaneous surface current
which flows within a depth of a few $\xi_0$.
The  two members of a pair of  time reversed states,
$d_{x^2-y^2}+ is$ and $d_{x^2-y^2}- is$ or  $d_{x^2-y^2}+ 
id_{xy}$  and  $d_{x^2-y^2}- id_{xy}$,
lead to  surface currents of opposite direction.
Typical results  for
the current density at a flat
surface, obtained from 
Fermi-liquid theory of superconductivity,
are shown in Fig.4.\\[-28pt]
\begin{figure}[t]
\vspace*{10mm}
\begin{center}
\begin{minipage}[b]{50mm}
\epsfysize=63mm {\epsfbox{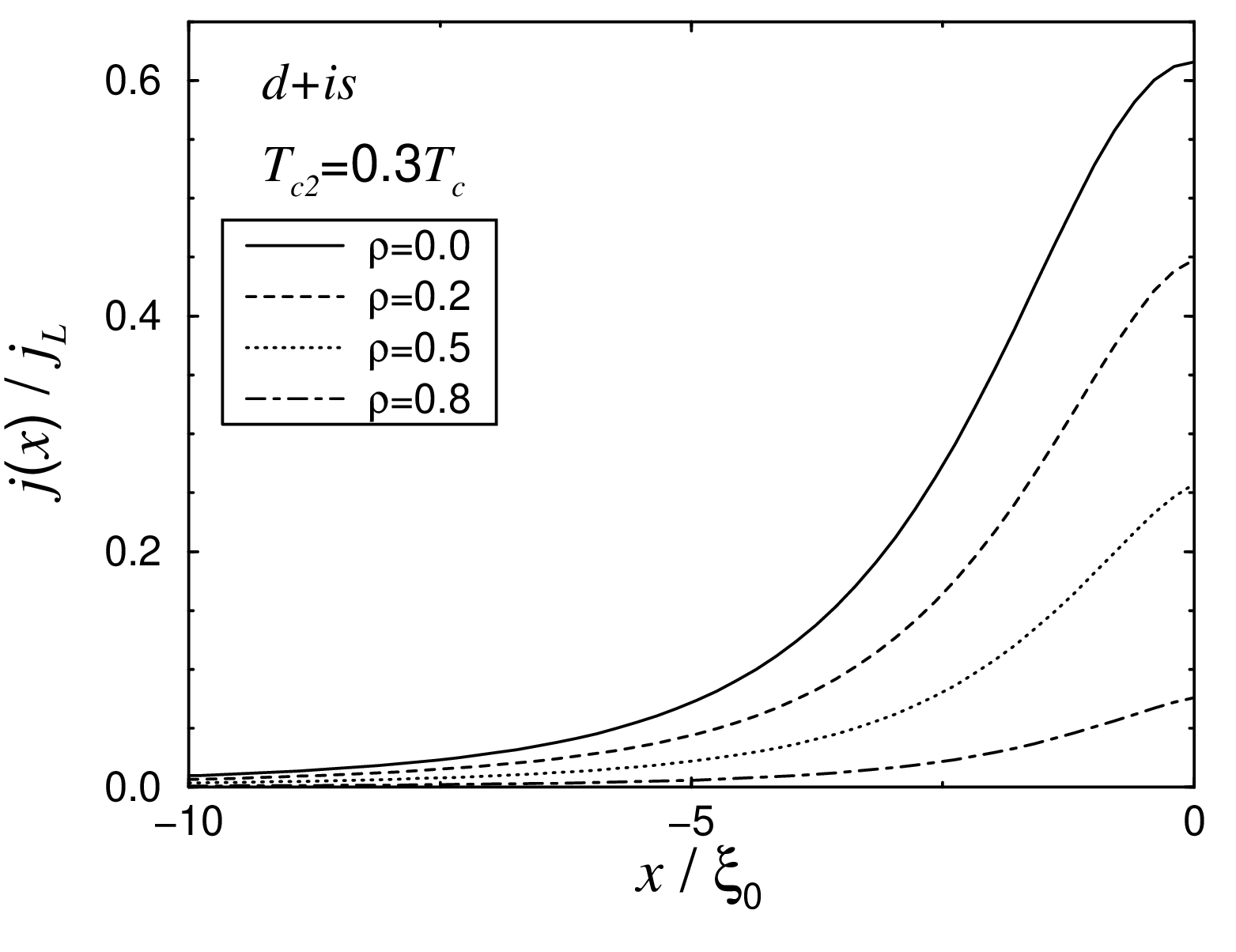}}
\end{minipage}\\
\begin{minipage}[b]{50mm}
\epsfysize=63mm {\epsfbox{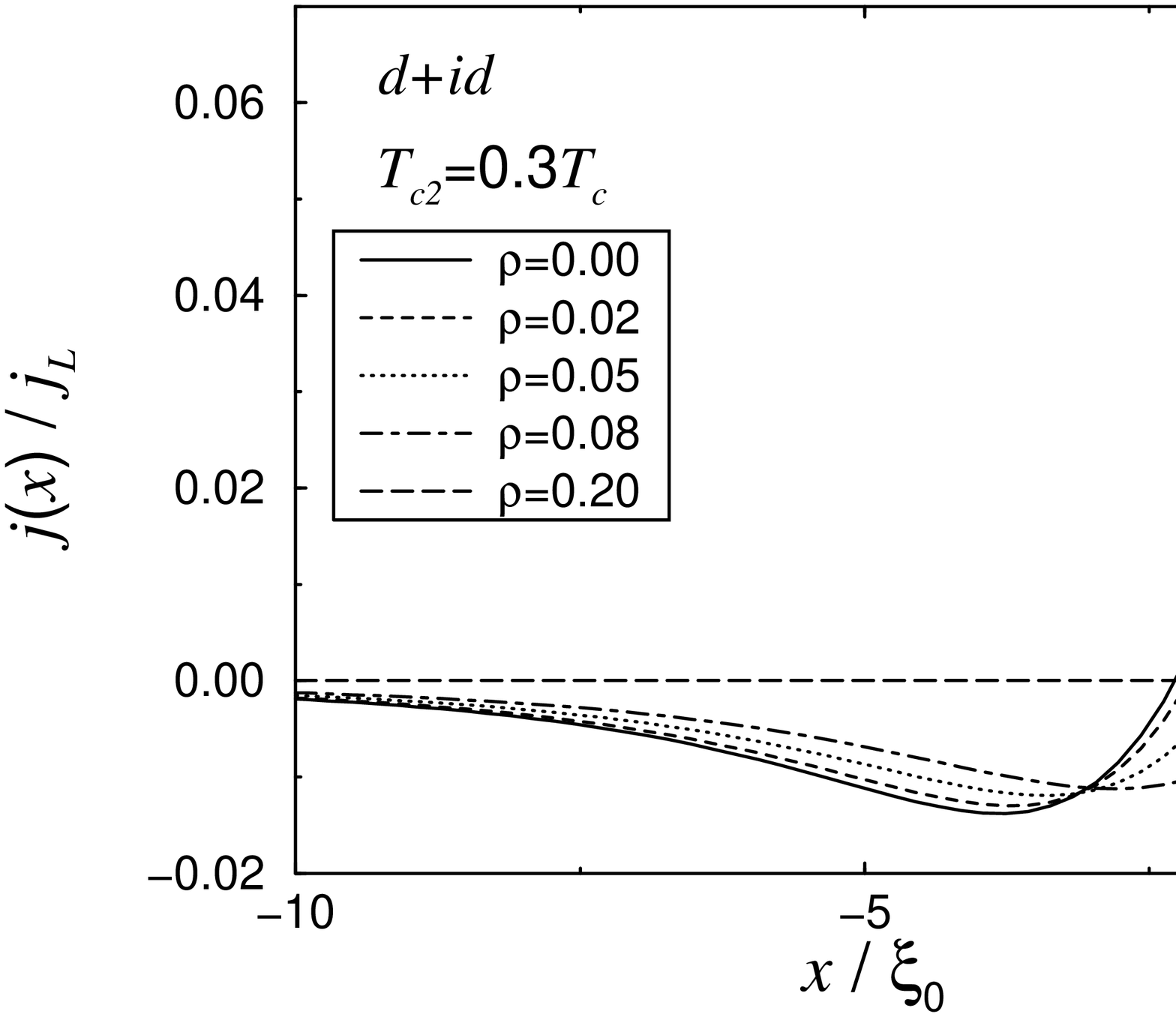}}
\end{minipage}
\end{center}
\caption{Current density (normalized by 
Landau's critical current, $\j_L$)
at a (110) surface
as function of the distance from the surface. 
Numerical results are presented  for 
$T_{c2}/T_{c}=0.3$,  $T/T_{c}=0.1$,  
various degrees of surface roughness ($\rho$),
and  two different 
subdominant order parameters.
Surface roughness is described by 
Ochchinnikov's model.
\label{Fig4}
}
\end{figure}
 
The current 
resulting from   a subdominant order parameter of type $s$
is an order of magnitude larger than from   $d_{xy}$, 
and much less sensitive to surface roughness. 
Important differences between surface states $d_{x^2-y^2}+is$ and
$d_{x^2-y^2}+id_{xy}$ are expected if the surface currents
have to be matched along a closed surface consisting of,
 say, a (110), (1-10), (-1-10)
and  (-110) surface. The differences 
follow from  the different symmetries of the two states.
For instance, the state  $d_{x^2-y^2}+ id_{xy}$ is 
invariant under  the combined
operation of time reversal and reflection
 $x\rightarrow -x$ (or $y\rightarrow -y$). As a consequence, the 
surface currents of the state $d_{x^2-y^2}+ id_{xy}$ can be matched  
smoothely in the four  corners to give a closed  current loop. 
On the other hand,
the state $d_{x^2-y^2}\pm is$ is 
invariant under the reflections $x\rightarrow -x$ 
and $y\rightarrow -y$, and the naive choice of 
a surface order parameter, namely 
the same state (e.g.  $d_{x^2-y^2}+ is$) 
on the (110), (1-10), (-1-10) and (-110)   surface
 leads to surface currents with sinks (sources) at the corners.
In order  to have  a single closed current loop for 
$d_{x^2-y^2}\pm is$-order, the surface states at
adjacent surfaces have to be time-reversed.
This requires    at each corner a  ``defect'' 
in the s-component at which $d_{x^2-y^2}\pm is$
changes into $d_{x^2-y^2}\mp is$. This defect covers an 
area $\approx\xi_0^2$ in the $x-y$ plane, and thus
costs little energy compared to the energy gained by forming 
the surface phases. 
The symmetry of the  stable surface phase 
is determined for macroscopic surfaces 
by the channel with maximum subdominant T$_c$. 
Measurements of
the surface transition temperature and  the symmetry
of the surface phase\cite{cov97} provide new 
insights into the pairing interaction and thus into 
the mechanism of superconductivity in  the materials of interest.

The work of J.A.S. was supported in part by the STC
for Superconductivity through NSF Grant no.\ 91-20000.
D.R. and J.A.S. also acknowledge support from the
Max-Planck-Gesellschaft and the Alexander von Humboldt-Stiftung.
M.F. acknowledges partial support from SF{\AA}AF,
{\AA}bo Akademi and M. Ehrnrooths Stiftelse.

\end{document}